\documentclass[11pt,a4paper]{article} 
\usepackage{a4wide}
\usepackage{amsmath}
\usepackage{graphicx}
\usepackage{makecell}
\usepackage{verbatim}
\usepackage{multirow}
\usepackage{subfig}
\usepackage{bm}
\usepackage{braket}
\usepackage{mathtools}
\usepackage{amssymb}
\usepackage{amscd}
\usepackage{latexsym}
\usepackage{slashed}
\usepackage[normalem]{ulem}
\usepackage{verbatim}
\usepackage{rotating}
\usepackage{diagbox}
\usepackage{cancel}
\usepackage[utf8]{inputenc}
\usepackage{array}
\usepackage{epsfig,amsfonts,amsthm}
\usepackage{float}
\usepackage{slashed}
\newlength{\absize}
\usepackage{verbatim}
\usepackage{cite}
\numberwithin{equation}{section}

\begin{document}

\thispagestyle{empty}
\renewcommand{\thefootnote}{\fnsymbol{footnote}}
\newpage\normalsize
\pagestyle{plain}
\setlength{\baselineskip}{4ex}\par
\setcounter{footnote}{0}
\renewcommand{\thefootnote}{\arabic{footnote}}
\renewcommand{\title}[1]{
\begin{center}
\LARGE #1
\end{center}\par}
\renewcommand{\author}[1]{
\vspace{2ex}
{\Large
\begin{center}
 \setlength{\baselineskip}{3ex} #1 \par
\end{center}}}
\renewcommand{\thanks}[1]{\footnote{#1}}
\renewcommand{\abstract}[1]{
\vspace{2ex}
\normalsize
\begin{center}
\centerline{\bf Abstract}\par
\vspace{2ex}
\parbox{\absize}{#1\setlength{\baselineskip}{3.5ex}\par}
\end{center}}

\vspace*{4mm}

\title{\textbf{Quark masses and mixing in the \boldmath{$D_5$} model under spontaneous CP violation} }

\author{Dong-Ping Fu$^{a}$\footnote{E-mail: fudongp@alumni.sysu.edu.cn} 
and 
Michihisa Takeuchi$^{a}$\footnote{E-mail: takeuchi@mail.sysu.edu.cn}}

\begin{center}
$^{a}$School of Physics and Astronomy, Sun Yat-sen University, 519082 Zhuhai, China 
\end{center}

\vfill

\abstract{ It is known that CP violation occurs in flavor physics, and the CKM matrix is complex. We investigate the Yukawa sector of a four-Higgs model based on $D_5$ symmetry. To understand all possible sources of CP violation in the Yukawa sector, we systematically analyze the impact of all possible representation assignments of the left-handed and right-handed quark fields under the $D_5$ group. After imposing the conditions of non-block-diagonal CKM matrix and the absence of massless quarks, we find that the only viable representation consists of one $D_5$ doublet and one singlet. In this case, two distinct types of quark mass models emerge in the quark sector. For each model, we introduce the most general vacuum that allows spontaneous CP violation (SCPV) into the mass matrices. Due to the specific structure of the $D_5$ symmetry, the vacuum expectation values (vevs) satisfy particular relations such that, for both quark models, the up- and down-quark mass squared matrices (i.e., $M_d M_d^\dagger$ and $M_u M_u^\dagger$) can be simultaneously block-diagonalized by the same unitary transformation, forcing the CKM matrix to be block-diagonal and thus yielding vanishing mixing angles and CP violation. This result rules out the possibility of explaining experimental observations within the $D_5$ 4HDM under the SCPV framework. }
\vspace*{30mm} \setcounter{footnote}{0} \vfill
\newpage
\renewcommand{\thefootnote}{\arabic{footnote}}

\setcounter{page}{1}
\section{Introduction}
\label{Sec:intro}

The Standard Model (SM) of particle physics is one of the most successful achievements of modern physics, but it is unlikely to be the ultimate theory because there are still many unexplained problems. First, the magnitude of CP violation provided by the Standard Model is far from sufficient to explain the observed matter-antimatter asymmetry in the Universe~\cite{Sakharov:1967dj,Kuzmin:1985mm}, and it also cannot explain the observed “hierarchy” of fermion masses and mixing~\cite{Cabibbo:1963yz,Gell-Mann:1964ewy,Glashow:1970gm,Kobayashi:1973fv,SLAC-SP-017:1974ind,Georgi:1974sy,E288:1977xhf}. In the field of new physics beyond the Standard Model, multi-Higgs models that extend the Higgs scalar sector~\cite{Branco:1985pf,Wu:1994ja,Lavoura:1994fv,Branco:2005em,Nishi:2006tg,Branco:2011iw,Inoue:2014nva,Barradas-Guevara:2015rea,deMedeirosVarzielas:2016rii,Grzadkowski:2016szj,Emmanuel-Costa:2016vej,Nierste:2019fbx,Nebot:2019qvr,Okada:2020brs,Kuncinas:2023ycz,Miro:2024zka,Gao:2024xte}, due to their possibility of providing additional sources of CP violation, have long been one of the key research directions. Meanwhile, in order to reduce the number of free parameters in generic multi-Higgs models, discrete symmetries are usually introduced~\cite{Altarelli:2005yx,Machado:2010uc,Keus:2013hya,Petcov:2018snn,Yao:2020qyy,Babu:2023oih,Abe:2023ilq}, which can also be used to explain various seemingly independent phenomena in flavor physics.

In Ref~\cite{Fu:2026agd}, we constructed a four-Higgs-doublet model with an exact $D_5$ symmetry ($D_5$ 4HDM) and presented the complete vacuum structure, including both CP-conserving and spontaneously CP‑violating cases. In general, provided that the constraints on the scalar potential parameters and the positive-definiteness condition of the Hessian are satisfied, these vacua can all be local minima of the potential. Each vacuum corresponds to a parameter space $S$ formed by the parameters $\lambda_i$, and within this parameter space the vacuum can potentially be the global minimum point. Usually, to verify whether these vacua are global minima, one needs to find the complete global minimum conditions. However, because the complete bounded-from-below (BFB) conditions for the $D_5$ 4HDM are not yet known, and there is a large variety of vacua, the research workload is enormous, making this task extremely difficult. To improve research efficiency, we adopt a more feasible strategy: we first assume these vacua are global minima, directly study their resulting effects in the Yukawa sector, select the vacua that are consistent with phenomenological experiments, and then investigate the global minimum conditions for these qualified vacua, thereby avoiding ineffective research.

In Ref~\cite{Fu:2026gws}, it has been demonstrated that, within the framework of explicit CP violation in the $D_5$ 4HDM, the most general CP-conserving vacuum R-N-4a can successfully fit the quark mass spectrum and the CKM matrix, and can also provide new sources of CP violation. In this paper, we carry out a more systematic study of the full quark sector of the $D_5$ 4HDM, focusing on the framework of spontaneous CP violation, and investigate whether its vacua can yield the experimentally required nonzero quark masses and a non-block-diagonal CKM matrix, as well as provide sources of CP violation.

The remainder of this paper is organized as follows. In Sec~\ref{Sec:D5}, we present the tensor product decomposition of the $D_5$ symmetry and the scalar potential. In Sec~\ref{Sec:quark}, we assign specific $D_5$ charges to the quark fields and discuss all possible group representation assignments involving the quark fields. We find that only when all quark fields transform as combinations of doublets and singlets of the $D_5$ group can the experimentally required nonzero quark masses and a non-block-diagonal CKM matrix be satisfied. In Sec~\ref{Sec:quarkscpv}, within the framework of spontaneous CP violation, we take the vacuum C-N-3a as an example and concretely study whether it can satisfy the experimentally required nonzero quark masses and a non-block-diagonal CKM matrix. Finally, Section~\ref{Sec:Conclusion} summarizes our conclusions. 

\section{The $D_5$ symmetry and the scalar potential}
\label{Sec:D5}
The group $D_{5}$ is the symmetry of a regular pentagon, and it has 10 elements divided into four irreducible representations, namely, $\textbf{1}, \textbf{1}'$ and two doublet: $\textbf{2}, \textbf{2}'$~\cite{Ishimori:2010au}. The multiplication rules are~\cite{Fu:2026agd}
\begin{equation}
\label{D511}
\begin{split}
  &\textbf{1} \otimes \textbf{1}= \textbf{1}' \otimes \textbf{1}'=\textbf{1},\quad 
  \textbf{1} \otimes \textbf{1}'=\textbf{1}'.\\
&\textbf{1} \otimes \textbf{2}= \textbf{1}' \otimes \textbf{2}=\textbf{2}, \quad \textbf{1} \otimes \textbf{2}'=\textbf{1}' \otimes \textbf{2}'=\textbf{2}',\\&  
  \textbf{2} \otimes \textbf{2}=\textbf{1}\oplus\textbf{1}'\oplus\textbf{2}', \quad \textbf{2}' \otimes \textbf{2}'=\textbf{1}\oplus\textbf{1}'\oplus\textbf{2}, \quad  \textbf{2} \otimes \textbf{2}'=\textbf{2}\oplus\textbf{2}'.\\
  \end{split}
\end{equation}

In $D_5$, the Clebsch-Gordan coefficients for all the tensor products are given as follows~\cite{Fu:2026agd}. The singlet $w$ with the singlet $z$ gives
\begin{equation}
\label{D511}
\begin{aligned}
&(w)_{\mathbf{1}}\otimes  (z)_{\mathbf{1}}=(w)_{\mathbf{1}'}\otimes (z)_{\mathbf{1}'} =(wz)_{\mathbf{1}}, \qquad (w)_{\mathbf{1}}\otimes  (z)_{\mathbf{1}'}=(wz)_{\mathbf{1}'}.
\end{aligned}
\end{equation}
The singlet $w$ with the doublet $ x=(x_1, x_2)^\intercal $, gives
\begin{equation}
\label{D5C12}
\begin{aligned}
&(w)_{\mathbf{1}}\otimes 
\begin{pmatrix}
x_1 \\
x_2 
\end{pmatrix}_{\bar{\mathbf{2}} }
=\begin{pmatrix}
wx_1 \\
wx_2 
\end{pmatrix}_{\bar{\mathbf{2}}},\qquad
(w)_{\mathbf{1}'}\otimes 
\begin{pmatrix}
x_1 \\
x_2 
\end{pmatrix}_{\bar{\mathbf{2}}}
=\begin{pmatrix}
wx_1 \\
-wx_2 
\end{pmatrix}_{\bar{\mathbf{2}}},
\end{aligned}
\end{equation}
where $\bar{\mathbf{2}}=\mathbf{2}, \mathbf{2}'$. The two doublets, $x=(x_1 , x_2)^\intercal $ and $ y=(y_1 , y_2)^\intercal $ , give
\begin{equation}
\label{D5C22}
\begin{aligned}
&\begin{pmatrix}
x_1 \\
x_2 
\end{pmatrix}_{\mathbf{2}}\otimes 
\begin{pmatrix}
y_1 \\
y_2 
\end{pmatrix}_{\mathbf{2}}
=(x_1y_2+x_2y_1)_{\mathbf{1}}
\oplus (x_1y_2-x_2y_1)_{\mathbf{1}'}
\oplus 
\begin{pmatrix}
x_1y_1 \\
x_2y_2 
\end{pmatrix}_{\mathbf{2}'}, \\&
\begin{pmatrix}
x_1 \\
x_2 
\end{pmatrix}_{\mathbf{2}'}\otimes 
\begin{pmatrix}
y_1 \\
y_2 
\end{pmatrix}_{\mathbf{2}'}
=(x_1y_2+x_2y_1)_{\mathbf{1}}
\oplus (x_1y_2-x_2y_1)_{\mathbf{1}'}
\oplus 
\begin{pmatrix}
x_2y_2 \\
x_1y_1 
\end{pmatrix}_{\mathbf{2}}, \\&
\begin{pmatrix}
x_1 \\
x_2 
\end{pmatrix}_{\mathbf{2}}\otimes 
\begin{pmatrix}
y_1 \\
y_2 
\end{pmatrix}_{\mathbf{2}'}
=\begin{pmatrix}
x_2y_1 \\
x_1y_2
\end{pmatrix}_{\mathbf{2}} 
\oplus 
\begin{pmatrix}
x_2y_2 \\
x_1y_1
\end{pmatrix}_{\mathbf{2}'}. 
\end{aligned}
\end{equation}

In the 4HDM, there are four scalar fields $\phi_1$, $\phi_2$, $\phi_3$, and $\phi_4$, which are all $SU(2)$ doublets. We denote the transformations of these four scalar fields under $D_5$ symmetry as
\begin{equation}
\label{Phi1234}
\begin{pmatrix}
\phi _1    \\
\phi _2    \\
\end{pmatrix}\sim \mathbf{2},\quad \quad 
\begin{pmatrix}
\phi _3    \\
\phi _4    \\
\end{pmatrix}\sim \mathbf{2}'.
\end{equation}
Thus, the most general scalar potential of $D_5$-symmetric 4HDM is given by~\cite{Fu:2026agd}
\begin{equation}
\begin{aligned}
\label{VD5}
V=&-\mu^2_{1}(\phi^\dagger_2\phi_2+\phi^\dagger_1\phi_1)-\mu^2_{2}(\phi^\dagger_4\phi_4+\phi^\dagger_3\phi_3)+\lambda_1\left[(\phi^\dagger_1\phi_1)^2+(\phi^\dagger_2\phi_2)^2\right]\\&+
\lambda_2\left[(\phi^\dagger_3\phi_3)^2+(\phi^\dagger_4\phi_4)^2\right]+\lambda_3(\phi^\dagger_1\phi_1)(\phi^\dagger_2 \phi_2)+\lambda_4(\phi^\dagger_3\phi_3)(\phi^\dagger_4 \phi_4)\\
&+\lambda_5\left[(\phi^\dagger_1\phi_1)(\phi^\dagger_3\phi_3)+(\phi^\dagger_2\phi_2)(\phi^\dagger_4\phi_4)\right]+\lambda_6\left[(\phi^\dagger_2\phi_2)(\phi^\dagger_3\phi_3)+(\phi^\dagger_1\phi_1)(\phi^\dagger_4\phi_4)\right]\\
&+\lambda_{7}(\phi^\dagger_1\phi_2)(\phi^\dagger_2\phi_1)
+\lambda_{8}(\phi^\dagger_3\phi_4)(\phi^\dagger_4\phi_3)+\lambda_{9}\left[(\phi^\dagger_1\phi_3)(\phi^\dagger_3\phi_1)+(\phi^\dagger_2
\phi_4)(\phi^\dagger_4\phi_2)\right]\\
&+\lambda_{10}\left[(\phi^\dagger_1\phi_4)(\phi^\dagger_4\phi_1)+(\phi^\dagger_2\phi_3)(\phi^\dagger_3\phi_2)\right] +\left\{ 
\lambda_{11}(\phi^\dagger_1\phi_3)(\phi^\dagger_2\phi_4)
+\lambda_{12}(\phi^\dagger_2\phi_3)(\phi^\dagger_1\phi_4)
\right.\\
&\left.+\lambda_{13}\left[(\phi^\dagger_2\phi_1)(\phi^\dagger_2\phi_3)+(\phi^\dagger_1\phi_2)(\phi^\dagger_1\phi_4)\right]
+\lambda_{14}\left[(\phi^\dagger_1\phi_3)(\phi^\dagger_4\phi_3)+(\phi^\dagger_2\phi_4)(\phi^\dagger_3\phi_4)\right]
+\text{h.c.} \right\}.
\end{aligned}
\end{equation}
The complete vacuum structure, including both CP-conserving and spontaneously CP-violating vacua, was also presented in that reference.
\section{Quark sector with $D_5$ symmetry}
\label{Sec:quark}
\subsection{Yukawa couplings in the quark sector}

In a general four-Higgs-doublet model (4HDM), the Yukawa couplings in the quark sector can be written as a sum over all four Higgs doublets, with the Yukawa Lagrangian given by \cite{Fu:2026gws}:
\begin{equation}
-\mathcal{L}_Y = \bar{q}_L \left( \sum_{i=1}^{4} \Gamma_i \phi_i \right) d'_R + \bar{q}_L \left( \sum_{i=1}^{4} \Delta_i \tilde{\phi}_i \right) u'_R +\text{h.c.},
\end{equation}
where $q_L = (u'_L, d'_L)^T$ is a column vector in the three-generation space, whose three components are the three generations of left-handed SU(2) quark doublet fields, and can be expressed as 
\begin{equation}
q_L^i= \begin{pmatrix} u'_{iL} \\ d'_{iL}\end{pmatrix}, \quad i=1,2,3.
\end{equation}
$u'_R$ ($d'_R$) is a vector in the three-dimensional right-handed up-type (down-type) quark space, with its components $u'_{iR}$ ($d'_{iR}$) being the $i$-th generation right-handed SU(2) up-type (down-type) quark singlet fields, i.e.,
\begin{equation} 
  u'_R = \begin{pmatrix} u_R \\ c_R \\ t_R \end{pmatrix}= \begin{pmatrix} u'_{1R} \\ u'_{2R} \\ u'_{3R}\end{pmatrix},\qquad
  d'_R = \begin{pmatrix} d_R \\ s_R \\ b_R \end{pmatrix}= \begin{pmatrix} d'_{1R} \\ d'_{2R} \\ d'_{3R} \end{pmatrix}.
\end{equation}
\(\tilde{\phi}_i = i\sigma_2 \phi_i^*\) is the conjugate representation of the Higgs doublet \(\phi_i\), and \(\Delta_i\) and \(\Gamma_i\) are \(3\times 3\) Yukawa coupling matrices that contain the Yukawa coupling coefficients; they are constant matrices determined by the symmetries of the model and encode the flavor mixing information, describing the coupling of the \(i\)-th Higgs field to the quark fields. After spontaneous symmetry breaking, the Higgs fields acquire vacuum expectation values \(\langle \phi_i \rangle = \frac{v_i}{\sqrt{2}}\), yielding the physical mass matrices for the down-type and up-type quarks as linear combinations of the Yukawa coupling matrices with the Higgs VEVs:
\begin{equation}
\begin{split}
 M_d = \frac{1}{\sqrt{2}} \sum_i \Gamma_i v_i, \qquad M_u = \frac{1}{\sqrt{2}}\sum_i \Delta_i v_i^*.
    \end{split}
\end{equation}
These matrices are generally not diagonal. We define \(U_\alpha\) (\(\alpha = d_L, d_R, u_L, u_R\)) as the matrices that rotate the quarks to the mass basis:
\begin{equation}
\begin{split}
  & \bar d'_L=\bar d_LU^\dagger_{d_L}, \qquad d'_R=U_{d_R}d_R,\\&
   \bar u'_L=\bar u_L U^\dagger_{u_L},
   \qquad u'_R=U_{u_R}u_R.
\end{split}
\end{equation}
The quark mixing CKM matrix is given by
\begin{equation}
\label{equ:CKM}
  V_{CKM}=U^\dagger_{u_L}U_{d_L}.
\end{equation}
The mass matrix \(M_d\) (\(M_u\)) is diagonalized by the unitary matrices \(U^\dagger_{d_L}\) and \(U_{d_R}\) (\(U^\dagger_{u_L}\) and \(U_{u_R}\)). We obtain the diagonalized mass matrices for the down-type and up-type quarks as follows:
\begin{equation}
\label{Dmd}
\begin{split}
&\mathcal{M}_d=\mathrm{diag}(m_d,m_s,m_b)=U^\dagger_{d_L} M_d U_{d_R}=\frac{1}{\sqrt{2}}U^\dagger_{d_L}\Bigl[\sum_i v_i\Gamma_i\Bigr]U_{d_R}, \\&
\mathcal{M}_u=\mathrm{diag}(m_u,m_c,m_t)=U^\dagger_{u_L} M_u U_{u_R}=\frac{1}{\sqrt{2}}U^\dagger_{u_L}\Bigl[\sum_i v_i\Delta_i\Bigr]U_{u_R}.
\end{split}
\end{equation}
After diagonalization, the mass squared matrices are
\begin{equation}
\begin{split}
&\mathcal{M}^2_d=\mathrm{diag}(m^2_d,m^2_s,m^2_b)=U^\dagger_{d_L}M_d M^\dagger_d U_{d_L},\\& \mathcal{M}^2_u=\mathrm{diag}(m^2_u,m^2_c,m^2_t)=U^\dagger_{u_L}M_u M^\dagger_u U_{u_L}.
   \end{split}
\end{equation}
The non-diagonalized mass squared matrices, i.e., the Hermitian matrices, are
\begin{equation}
\begin{split}
&H_d=M_dM^\dagger_d=U_{d_L}\,\mathrm{diag}(m^2_d,m^2_s,m^2_b)\,U^\dagger_{d_L},\\&H_u= M_uM^\dagger_u=U_{u_L}\,\mathrm{diag}(m^2_u,m^2_c,m^2_t)\,U^\dagger_{u_L}.
   \end{split}
\end{equation}

\subsection{Representations of quark fields under the $D_5$ symmetry}
\indent Since the $D_5$ group has two singlet representations: $\textbf{1}$, $\textbf{1}'$, and two doublet representations: $\textbf{2}$, $\textbf{2}'$, there are many different possible representation assignments for the scalar, quark, and lepton fields. Since permutations of the three fields within each sector do not lead to new structures of the quark mass matrices, we use “$3\textbf{s}$” to denote that each of the three generations of quark fields transforms as a singlet representation, and then there are the following independent possibilities:
\begin{equation}
\label{D5111}
\begin{split}
  &(\textbf{1},\textbf{1},\textbf{1}),\quad
  (\textbf{1}',\textbf{1}',\textbf{1}'), \quad
  (\textbf{1},\textbf{1}',\textbf{1}'), \quad
  (\textbf{1}',\textbf{1},\textbf{1}). 
  \end{split}
\end{equation}
Similarly, we use “$\textbf{d},\textbf{s}$” to denote that among the three generations of quark fields, two generations form a doublet representation and the remaining one forms a singlet representation, and then there are the following independent possibilities:
\begin{equation}
\label{D52plus1}
\begin{split}
  &(\textbf{2},\textbf{1}),\quad
  (\textbf{2},\textbf{1}'), \quad
  (\textbf{2}',\textbf{1}), \quad
  (\textbf{2}',\textbf{1}'). 
  \end{split}
\end{equation}
Since the Yukawa Lagrangian must transform as the invariant singlet representation $\textbf{1}$ in order for the corresponding matrix elements of the Yukawa coupling matrices to be non-zero, and in our $D_5$ 4HDM the four Higgs doublets are in the representations $\textbf{2}$ and $\textbf{2}'$, we know from the tensor product decomposition formula of the $D_5$ group~(\ref{D511}) that the product of the left-handed SU(2) quark doublet ($q_L$) and the right-handed quark SU(2) singlets (up-type $u'_R$ or down-type $d'_R$) must also be in a doublet representation. This can only be realized if at least one of the left- or right-handed quark fields is in a doublet representation. We list five possible representation assignments in Table~\ref{Table:yukawa2plus1}.
\begin{table}[H]
\centering
\renewcommand{\arraystretch}{1.5}
\begin{tabular}{cccc}
\hline 
Type &$\bar q_L$ &   $d'_R$        & $u'_R$           \\
\hline 
I&$\textbf{d},\textbf{s}$&$ \textbf{d},\textbf{s} $&$\textbf{d},\textbf{s}$\\
II&$\textbf{d},\textbf{s}$ & $ 3\textbf{s}$ & $\textbf{d},\textbf{s}$ \\
III&$\textbf{d},\textbf{s}$ & $\textbf{d},\textbf{s}$ &$3\textbf{s}$  \\
IV&$\textbf{d},\textbf{s}$ & $ 3\textbf{s}$ & $ 3\textbf{s}$ \\
V&$ 3\textbf{s}$ & $\textbf{d},\textbf{s}$ & $\textbf{d},\textbf{s}$ \\
\hline 
\end{tabular}
\caption{Possible representations for the left-handed quark $SU(2)_L$ doublet ($q_L$) and the right-handed quark $SU(2)_L$ singlets ($d_R'$ and $u_R'$).\label{Table:yukawa2plus1}}
\end{table}

\subsubsection{Type I: Both quark fields are a combination of doublet and singlet}

Now we consider the assignment where both left- and right-handed quark fields transform as combinations of doublet and singlet representations of the $D_5$ group, and first assume the following specific group representation assignment:
\begin{align}
\label{qnp222}
\textbf{Case A}\quad\quad \bar q_{L}\sim (\textbf{2}, \textbf{1}), \quad  d'_R \sim (\textbf{2}, \textbf{1}), \quad  u'_R \sim (\textbf{2}, \textbf{1}).
\end{align}
Then the Yukawa Lagrangian for the quark sector that preserves the gauge symmetry and the $D_5$ symmetry is:
\begin{subequations}
\begin{equation}
\label{caseAlyd}
\begin{aligned}
\mathcal{-L}_{Y_d}=\;&y_{1}^d(\bar q^1_{L}{\phi}_2 + \bar q^2_{L}{\phi} _1)d'_{3R} + y_{2}^d(\bar q^2_{L}{\phi}_3 d'_{2R} + \bar q^1_{L}{\phi}_4 d'_{1R})\quad \\& + y_{3}^d \bar q^3_{L}({\phi}_1 d'_{2R} + {\phi}_2 d'_{1R}) + \text{h.c.},
\end{aligned}
\end{equation} 
\begin{equation}
\begin{aligned}
\mathcal{-L}_{Y_u}=\;&y_{1}^u(\bar q^1_{L}\tilde{\phi}_1 + \bar q^2_{L}\tilde{\phi}_2)u'_{3R} + y_{2}^u(\bar q^2_{L}\tilde{\phi}_4 u'_{2R} + \bar q^1_{L}\tilde{\phi}_3 u'_{1R})\\& + y_{3}^u \bar q^3_{L}(\tilde{\phi}_2 u'_{2R} + \tilde{\phi}_1 u'_{1R}) + \text{h.c.},
\end{aligned}
\end{equation} 
\end{subequations}
Here $y_{i}^d$ ($i=1,2,3$) and $y_{i}^u$ ($i=1,2,3$) are the complex Yukawa coupling coefficients for the down-type and up-type quarks, respectively, with each containing three independent coefficients. Within the framework of SCPV, the Lagrangian is required to be explicitly CP-conserving, so these coupling coefficients are taken to be real parameters. In the framework of explicit CP violation, on the other hand, these coupling coefficients are taken to be complex parameters. $\text{h.c.}$ denotes the Hermitian conjugate terms, and we have used the standard abbreviation $\tilde{\phi}_i = i\sigma_2 \phi_i^*$. Note that for the up-type quarks, the scalar doublet $(\phi_1, \phi_2)^\intercal$ is replaced by $(\tilde{\phi}_2,\tilde{\phi}_1)^\intercal$, and $(\phi_3, \phi_4)^\intercal$ is replaced by $(\tilde{\phi}_4,\tilde{\phi}_3)^\intercal$. After spontaneous symmetry breaking, we obtain the mass matrices for the down-type and up-type quarks as follows:
\begin{subequations}
\label{caseAmassd}
\begin{equation}
\label{caseAmass}
M_{d}=\frac{1}{\sqrt{2}}\begin{pmatrix}
 y_{2}^du_4& 0&y_{1}^du_2\\
0&y_{2}^du_3& y_{1}^du_1\\
y_{3}^du_2& y_{3}^du_1& 0
\end{pmatrix},
\end{equation}
\begin{equation}
\label{caseAmassu}
M_{u}=\frac{1}{\sqrt{2}}\begin{pmatrix}
 y_{2}^uu^*_3& 0&y_{1}^uu^*_1\\
0&y_{2}^uu^*_4& y_{1}^uu^*_2\\
y_{3}^uu^*_1& y_{3}^uu^*_2& 0
\end{pmatrix}.
\end{equation}
\end{subequations}
The diagonalized quark mass matrices $\mathcal{M}_d$ ($\mathcal{M}_u$) can be obtained by diagonalizing the mass matrices $M_d$ ($M_u$) with two unitary matrices $U^\dagger_{d_L}$ and $U_{d_R}$ ($U^\dagger_{u_L}$ and $U_{u_R}$) according to the definition in Eq.~(\ref{Dmd}). The matrices $U_{d_L}$ and $U_{u_L}$ can be obtained by diagonalizing the mass-squared matrices $M_{d}M^\dagger_d$ and $M_{u}M^\dagger_u$, respectively. Then,
\begin{subequations}
\begin{equation}
\label{caseAdd}
M_{d}M_{d}^\dagger = \frac{1}{2} 
\begin{pmatrix}
|y_{1}|^2 |u_2|^2+|y_{2}|^2 |u_4|^2  & |y_{1}|^2 u_2 u_1^* & y_{2} y_{3}^* u_4 u_2^* \\
|y_{1}|^2 u_1 u_2^* & |y_{1}|^2 |u_1|^2+|y_{2}|^2 |u_3|^2  & y_{2} y_{3}^* u_3 u_1^* \\
y_{3} y_{2}^* u_2 u_4^* & y_{3} y_{2}^* u_1 u_3^* & |y_{3}|^2 (|u_1|^2+|u_2|^2)
\end{pmatrix},
\end{equation}
\begin{equation}
\label{caseAuu}
M_u M_u^\dagger = \frac{1}{2} 
\begin{pmatrix}
|y_{1}|^2 |u_1|^2+|y_{2}|^2 |u_3|^2  & |y_{1}|^2 u_2 u_1^*  & y_{2} y_{3}^* u_1 u_3^*  \\
|y_{1}|^2 u_1 u_2^* & |y_{1}|^2 |u_2|^2+|y_{2}|^2 |u_4|^2  & y_{2} y_{3}^* u_2 u_4^*  \\
y_{3} y_{2}^* u_3  u_1^*  & y_{3} y_{2}^* u_4 u_2^*  & |y_{3}|^2 (|u_1|^2+|u_2|^2)
\end{pmatrix}.
\end{equation}  
\end{subequations}
Note that, for notational simplicity, the superscripts for down-type and up-type Yukawa couplings have been omitted in the above expressions, i.e., $y_i = y_{i}^{(d,u)}$. The following text is the same. By solving the eigenvalue equations of these matrices, one obtains the eigenvalues; the three eigenvalues of $M_{d}M^\dagger_d$ are the squared masses of the three down-type quarks, i.e., $m^2_d,m^2_s,m^2_b$, while the three eigenvalues of $M_{u}M^\dagger_u$ are the squared masses of the three up-type quarks, i.e., $m^2_u,m^2_c,m^2_t$. The above mass-squared matrices (\ref{caseAdd}) and (\ref{caseAuu}) themselves do not lead to vanishing quark masses, nor do they make the CKM mixing matrix block-diagonal. However, in specific vacua, further analysis of the quark masses and CKM mixing is still required.\\
\indent Now consider the second group representation assignment for the quark fields transforming under the $D_5$ symmetry:
\begin{align}
\label{qnp22'2'}
\textbf{Case B}\quad\quad\bar q_{L}\sim (\textbf{2}, \textbf{1}), \quad  d'_R \sim (\textbf{2}', \textbf{1}), \quad  u'_R \sim (\textbf{2}', \textbf{1}).
\end{align}
In this case, the Yukawa Lagrangian for the quark sector that preserves the gauge symmetry and the $D_5$ symmetry is:
\begin{subequations}
\begin{equation}
\begin{aligned}
\mathcal{-L}_{Y_d}=\;&y_{1}^d(\bar q^1_{L}{\phi}_1 d'_{2R} + \bar q^2_{L}{\phi}_2 d'_{1R})+y_{2}^d(\bar q^1_{L}{\phi}_2 + \bar q^2_{L}{\phi} _1)d'_{3R}\\&+y_{3}^d(\bar q^2_{L}{\phi}_4 d'_{2R} + \bar q^1_{L}{\phi}_3 d'_{1R})+ y_{4}^d \bar q^3_{L}({\phi}_3 d'_{2R} + {\phi}_4 d'_{1R}) + \text{h.c.},
\end{aligned}
\end{equation}
\begin{equation}
\label{caseBlyu}
\begin{aligned}
\mathcal{-L}_{Y_u}=\;&y_{1}^u(\bar q^1_{L}\tilde{\phi}_2 u'_{2R} + \bar q^2_{L}\tilde{\phi}_1 u'_{1R})+y_{2}^u(\bar q^1_{L}\tilde{\phi}_1 + \bar q^2_{L}{\phi}_2)u'_{3R} \\&+y_{3}^u(\bar q^2_{L}\tilde{\phi}_3 u'_{2R} + \bar q^1_{L}\tilde{\phi}_4 u'_{1R})+ y_{4}^u \bar q^3_{L}(\tilde{\phi}_4 u'_{2R} + \tilde{\phi}_3 u'_{1R}) + \text{h.c.},
\end{aligned}
\end{equation} 
\end{subequations}
Compared with \textbf{Case A}, in this case each type of quark has four independent Yukawa coupling coefficients. After spontaneous symmetry breaking, the mass matrices for the down-type and up-type quarks are as follows:
\begin{subequations}
\begin{equation}
\label{caseBmassd}
M_{d}=\frac{1}{\sqrt{2}}\begin{pmatrix}
 y_{3}^du_3&  y_{1}^du_1&y_{2}^du_2\\
y_{1}^du_2&y_{3}^du_4& y_{2}^du_1\\
y_{4}^du_4& y_{4}^du_3& 0
\end{pmatrix},
\end{equation}
\begin{equation}
\label{caseBmassu}
M_{u}=\frac{1}{\sqrt{2}}\begin{pmatrix}
 y_{3}^uu^*_4&  y_{1}^uu^*_2&y_{2}^uu^*_1\\
 y_{1}^uu^*_1&y_{3}^uu^*_3& y_{2}^uu^*_2\\
y_{4}^uu^*_3& y_{4}^uu^*_4& 0
\end{pmatrix}.
\end{equation}
\end{subequations}

Then the mass-squared matrices for the down-type and up-type quarks are respectively:
\begin{subequations}
\begin{equation}
\begin{aligned}
\label{caseBdd}
&M_d M_d^\dagger = \\&\frac{1}{2}
\scalebox{0.92}{$\begin{pmatrix}
|y_1|^2|u_1|^2 + |y_2|^2|u_2|^2+|y_3|^2|u_3|^2 & y_3 y_1^* u_3 u_2^* + y_1 y_3^* u_1 u_4^* + |y_2|^2 u_2 u_1^* & y_3 y_4^* u_3 u_4^* + y_1 y_4^* u_1 u_3^* \\
y_1 y_3^* u_2 u_3^* + y_3 y_1^* u_4 u_1^* + |y_2|^2 u_1 u_2^* & |y_1|^2|u_2|^2  + |y_2|^2|u_1|^2 + |y_3|^2|u_4|^2 & y_1 y_4^* u_2 u_4^* + y_3 y_4^* u_4 u_3^* \\
y_4 y_3^* u_4 u_3^* + y_4 y_1^* u_3 u_1^* & y_4 y_1^* u_4 u_2^* + y_4 y_3^* u_3 u_4^* & |y_4|^2 (|u_3|^2 + |u_4|^2)
\end{pmatrix},$}
\end{aligned}
\end{equation}
\begin{equation}
\begin{aligned}
\label{caseBuu}
&M_u M_u^\dagger = \\&\frac{1}{2}
\scalebox{0.92}{$\begin{pmatrix}
|y_1|^2|u_2|^2+|y_2|^2|u_1|^2 +|y_3|^2|u_4|^2& y_3 y_1^* u_1 u_4^* + y_1 y_3^* u_3 u_2^*  + |y_2|^2 u_2 u_1^*  & y_1 y_4^* u_4 u_2^* + y_3 y_4^* u_3 u_4^*   \\
y_1 y_3^* u_4 u_1^*  + y_3 y_1^* u_2 u_3^* + |y_2|^2 u_1 u_2^*  & |y_1|^2|u_1|^2+|y_2|^2|u_2|^2 +|y_3|^2|u_3|^2& y_1 y_4^* u_3 u_1^*  + y_3 y_4^* u_4 u_3^*  \\
y_4 y_1^* u_2 u_4^* + y_4 y_3^* u_4 u_3^*   & y_4 y_1^* u_1 u_3^* + y_4 y_3^* u_3 u_4^* & |y_4|^2(|u_3|^2+|u_4|^2)
\end{pmatrix},$}
\end{aligned}
\end{equation}   
\end{subequations}
Similar to \textbf{Case A}, the above mass-squared matrices (\ref{caseBdd}) and (\ref{caseBuu}) do not lead to vanishing quark masses, nor do they make the CKM mixing matrix block-diagonal. However, in specific vacua, further analysis of the quark masses and CKM mixing is still required.

Among the assignments where all quark fields transform as combinations of doublets and singlets (see Eq.~(\ref{D52plus1})), there are four independent basic combinations. Through permutations, there are theoretically $64$ cases. However, in fact, by calculating other combinations from Eq.~(\ref{D52plus1}), for example, replacing the left- or right-handed fields in \textbf{Case A} with other doublet-singlet combinations, one finds that the eigenvalues (physical observables) of the new quark mass-squared matrices are exactly the same as those of the \textbf{Case A} or \textbf{Case B} mass-squared matrices, i.e., they do not lead to essentially new physical results. Therefore, in $D_5$ 4HDM, it is sufficient to consider only the quark mass models of \textbf{Case A} and \textbf{Case B} as representatives of all quark mass matrix structures that satisfy the experimental constraints, and there is no need to repeatedly consider other combinations that differ only by permutations of group representation assignments without yielding new physical predictions.

Another point worth emphasizing is that in both \textbf{Case A} and \textbf{Case B}, we have chosen to assign the same group representation transformations to the down-type and up-type quarks. One could of course assume that they transform under different group representations, but in that case the down-type quark mass-squared matrix would still take the form of Eq.~(\ref{caseAdd}) or (\ref{caseBdd}), while the up-type quark mass-squared matrix would still take the form of Eq.~(\ref{caseAuu}) or (\ref{caseBuu}). The only difference lies in the permutations arising from different representation assignments.\\
\indent For example, we can consider the following group representation assignment for the quark fields under the $D_5$ symmetry:
\begin{align}
\label{qnp222'}
\textbf{Case C}\quad\quad\bar q_{L}\sim (\textbf{2}, \textbf{1}), \quad  d'_R \sim (\textbf{2}, \textbf{1}), \quad  u'_R \sim (\textbf{2}', \textbf{1}).
\end{align}
In this case, the Yukawa Lagrangians for the up-type and down-type quarks that preserve the gauge symmetry and the $D_5$ symmetry are given by Eq.~(\ref{caseAlyd}) and Eq.~(\ref{caseBlyu}), respectively. After spontaneous symmetry breaking, the mass matrices for the down-type and up-type quarks are given by Eq.~(\ref{caseAmassd}) and Eq.~(\ref{caseBmassu}), respectively. Then the mass-squared matrices for the down-type and up-type quarks are given by Eq.~(\ref{caseAdd}) and Eq.~(\ref{caseBuu}), respectively.

\subsubsection{Type II: Only right-handed down-type quark fields are three singlets}

Consider the case where the right-handed down-type quark fields are taken as three singlet representations, while the other fields are combinations of doublet and singlet representations, and assume the following specific group representation assignment:
\begin{equation}
\label{typeii}
\bar q_{L}\sim (\textbf{2}, \textbf{1}), \qquad  d'_R \sim (\textbf{1},\textbf{1},\textbf{1}),  \qquad  u'_R \sim (\textbf{2}, \textbf{1}).
\end{equation}
After spontaneous symmetry breaking, we obtain the mass matrix for the down-type quarks as follows:
\begin{equation}
M_{d}=\frac{1}{\sqrt{2}}\begin{pmatrix}
 y_{1}^du_2&  y_{2}^du_2&y_{3}^du_2\\
y_{1}^du_1&y_{2}^du_1& y_{3}^du_1\\
0& 0& 0
\end{pmatrix}.
\end{equation}
One can see that the last row of the matrix is entirely zero. This is because the Higgs doublets $(\phi_1,\phi_2)^\intercal$ and $(\phi_3,\phi_4)^\intercal$ both belong to doublet representations of the $D_5$ symmetry group, so the product of the left-handed and right-handed quark fields must also be a doublet. However, $\bar q_{L}^3$ and all $d'_{iR}$ are singlets, which causes the non-zero elements of the mass squared matrix $M_d M_d^\dagger$ to appear only in the sub-block formed by the first two rows and first two columns, while the third row and third column are entirely zero. That is,
\begin{equation}
M_{d}M_{d}^\dagger=\frac{1}{2} \left(|y_{1}^d|^2+|y_{2}^d|^2+|y_{3}^d|^2\right)
\begin{pmatrix}
|u_2|^2 & u_2 u_1^* & 0 \\
u_1 u_2^* & |u_1|^2 & 0 \\
0 & 0 & 0
\end{pmatrix}.
\end{equation}
In this case, two massless down-type quarks would appear; however, all experimentally observed quark masses are non-zero. Therefore, the Type II quark field group representation, as exemplified by the assignment in Eq.~(\ref{typeii}), yields mass matrices inconsistent with experimental features and should be excluded in the $D_5$ 4HDM.

\subsubsection{Type III: Only right-handed up-type quark fields are three singlets}

Consider the case where the right-handed up-type quark fields are taken as three singlet representations, while the other fields are combinations of doublet and singlet representations, and assume the following specific group representation assignment:
\begin{equation}
\label{typeiii}
\bar q_{L}\sim (\textbf{2}, \textbf{1}), \qquad  d'_R \sim (\textbf{2}, \textbf{1}),  \qquad  u'_R \sim (\textbf{1},\textbf{1},\textbf{1}).
\end{equation}
After spontaneous symmetry breaking, we obtain the mass matrix for the up-type quarks as follows:
\begin{equation}
M_{u}=\frac{1}{\sqrt{2}}\begin{pmatrix}
 y_{1}^uu_1^*&  y_{2}^uu_1^*&y_{3}^uu_1^*\\
y_{1}^uu_2^*&y_{2}^uu_2^*& y_{3}^uu_2^*\\
0& 0& 0
\end{pmatrix}.
\end{equation}
Similar to Type II, here the non-zero elements of the mass squared matrix $M_u M_u^\dagger$ appear only in the sub-block formed by the first two rows and first two columns, while the third row and third column are entirely zero, leading to two massless up-type quarks. Therefore, the Type III quark field group representation, as exemplified by the assignment in Eq.~(\ref{typeiii}), should also be excluded in the $D_5$ 4HDM.

\subsubsection{Type IV: Both right-handed quark fields are three singlets}

Consider the case where both the right-handed up-type and down-type quark fields are taken as three singlets, and the left-handed quark fields are a combination of doublets and singlets, i.e.,
\begin{align}
\bar q_{L}\sim (\textbf{d}, \textbf{s}), \qquad  d'_R \sim 3\textbf{s} ,  \qquad  u'_R \sim 3\textbf{s}.
\end{align}
Combining Type II and Type III, we see that this case leads simultaneously to two massless up-type quarks and two massless down-type quarks. Therefore, the Type IV quark field group representation assignment should likewise be excluded in the $D_5$ 4HDM.

\subsubsection{Type V : Only left-handed quark fields are three singlets}

Consider the case where the left-handed quark fields are taken as three singlet representations, while the right-handed quark fields are combinations of doublet and singlet representations, and assume the following specific group representation assignment:
\begin{align}
\label{typeV}
\bar q_{L}\sim \textbf{s}, \qquad  d'_R \sim (\textbf{2}, \textbf{1}) ,  \qquad  u'_R \sim (\textbf{2}, \textbf{1}).
\end{align}
We obtain the mass matrices for the down-type and up-type quarks, respectively, as follows:
\begin{equation}
M_{d}=\frac{1}{\sqrt{2}}\begin{pmatrix}
 y_{1}^du_2&  y_{1}^du_1&0\\
y_{2}^du_2&y_{2}^du_1& 0\\
y_{3}^du_2& y_{3}^du_1& 0
\end{pmatrix},\qquad M_{u}=\frac{1}{\sqrt{2}}\begin{pmatrix}
 y_{1}^uu_1^*&  y_{1}^uu_2^*&0\\
y_{2}^uu_1^*&y_{2}^uu_2^*& 0\\
y_{3}^uu_1^*& y_{3}^uu_2^*& 0
\end{pmatrix}.
\end{equation}
One can see that the last column of both matrices is entirely zero. This is because the Higgs doublets $(\phi_1,\phi_2)^\intercal$ and $(\phi_3,\phi_4)^\intercal$ both belong to doublets of the $D_5$ symmetry group, so the product of the left-handed and right-handed quark fields must also be a doublet. However, $\bar q_{L}^i$ and all $d'_R$ are singlets. Then the non-zero elements of the mass squared matrices $M_d^\dagger M_d$ and $M_u^\dagger M_u$ appear only in the sub-blocks formed by the first two rows and first two columns, while the third row and third column are entirely zero, which necessarily leads to two massless down-type quarks and two massless up-type quarks. Therefore, the Type V quark field group representation, as exemplified by the assignment in Eq.~(\ref{typeV}), should also be excluded in the $D_5$ 4HDM.

In summary, we have considered all possible group representation assignments for the quark fields under the $D_5$ symmetry group listed in Table~\ref{Table:yukawa2plus1}, and we conclude that only Type I, i.e., the assignment where both left- and right-handed quark fields are combinations of doublet and singlet representations, can ensure that the quark mass matrices of the model satisfy the experimental requirements that all quark masses are non-zero and the CKM matrix is non-block-diagonal. Moreover, all possible quark mass structures can be unified into the quark mass models of \textbf{Case A} and \textbf{Case B}, which are sufficient to cover all possible quark mass-squared matrices.

\section{Quark masses and mixing in SCPV}
\label{Sec:quarkscpv}
Within the framework of spontaneous CP violation in the $D_5$ 4HDM, we examine whether its vacua can satisfy the experimentally required non-zero quark masses and a non-block-diagonal CKM matrix. The complex vacua that can lead to SCPV, along with the phase conditions they must satisfy, are summarized in Table~\ref{scpvcvevs}~\cite{Fu:2026agd}.
\begin{table}[htb]
\centering
\renewcommand{\arraystretch}{1.3}
\begin{tabular}{|l|c|c|}
\hline 
Vacuum &Complex vevs &Constraint on phase\\
\hline 
C-N-3a & $v_1e^{i\theta},\pm v_1e^{i(\theta+\frac{n\pi}{5})}, \pm v_3e^{\frac{3in\pi}{5}},v_3$ & $\theta\neq\frac{n\pi}{5}$, \quad$ n\text{ integer}$, \\
\hline 
C-N-3b & $v_1e^{i\theta},\pm v_1e^{i\theta},\pm v_3,v_3$ &$\theta\neq k\pi $,\quad $k \text{ integer}$,\\
\hline 
C-N-3c&$v_1e^{i(\frac{1}{2}+\frac{n}{5})\pi},\pm v_1e^{i(\frac{1}{2}+\frac{2n}{5})\pi}, \pm v_3e^{\frac{3in\pi}{5}},v_3$   &--\\
\hline 
C-N-3d&$iv_1, \pm iv_1,\pm v_3,v_3$   &--\\
\hline 
\end{tabular}
\caption{Complex vacua with SCPV. “--” indicates no constraint on the phase.}
\label{scpvcvevs}
\end{table}

\subsection{C-N-3a Model I}

\indent We consider the most general complex vacuum C-N-3a listed in Table~\ref{scpvcvevs}, which has two forms: $(v_1e^{i\theta}, v_1e^{i(\theta+\frac{n\pi}{5})}, v_3e^{\frac{3in\pi}{5}},v_3)$ and $(v_1e^{i\theta},- v_1e^{i(\theta+\frac{n\pi}{5})}, - v_3e^{\frac{3in\pi}{5}},v_3)$. The constraints on the scalar potential parameters that they satisfy have been given in~\cite{Fu:2026agd}, with $v_1 \neq 0$ and $v_3 \neq 0$. These two vacua can be transformed into each other by applying an overall phase transformation to the second and third Higgs doublets: \(\phi_2 \to -\phi_2\), \(\phi_3 \to -\phi_3\) (i.e., multiplying by a phase of \(\pi\)).
Substituting these two sets of vacua into the scalar potential~(\ref{VD5}), one finds that they differ only by the sign of one potential parameter, $\lambda_{13}$, while all other parameters and all term forms remain exactly the same. This is precisely the direct manifestation of the above field redefinition on the scalar potential: under the transformation $\phi_2,\phi_3\to -\phi_2,-\phi_3$, the monomial corresponding to that parameter changes sign, while the remaining monomials remain unchanged. By redefining that parameter as $\lambda_{13}\to -\lambda_{13}$, the transformed scalar potential completely recovers its original form. Therefore, the scalar potential functions corresponding to these two sets of vacua are equivalent under parameter redefinition, and they describe the same physical vacuum. The differences among various vacuum expectation values are essentially relative magnitudes and relative phases, which are equivalent to relabeling the corresponding representation basis and do not break the symmetry form of the Lagrangian. Meanwhile, this operation can be compensated by adjusting the phases of the fermion fields that couple to $\phi_2$ and $\phi_3$, thereby keeping the Yukawa interactions unchanged. Since all observable physical quantities (scalar mass spectra, mixing angles, CP phases, coupling constants) are determined by the equivalence classes of the scalar potential function and the vacuum expectation values under parameter redefinition, the two sets of vacua must yield completely identical physical predictions. Furthermore, for other vacua, such as the two forms of C-N-3b: $(v_1e^{i\theta}, v_1e^{i\theta}, v_3, v_3)$ and $(v_1e^{i\theta}, -v_1e^{i\theta}, -v_3, v_3)$, the former retains a residual $Z_2$ symmetry while the latter appears to completely break the $D_5$ symmetry. However, this difference is merely an artifact of the basis choice and does not affect the physical results. Hence, in practical studies one only needs to choose one of the vacuum forms for analysis.\\
\indent We choose to study the vacuum $(v_1e^{i\theta}, v_1e^{i(\theta+\frac{n\pi}{5})}, v_3e^{\frac{3in\pi}{5}}, v_3)$, whose dependence on the integer $n$ enters only through the factors $e^{\frac{in\pi}{5}}$ and $e^{\frac{3in\pi}{5}}$. Since both factors have period $10$, the form of the vacuum is entirely determined by the residue of $n$ modulo $10$ (i.e., $n \bmod 10$). Hence there exist exactly ten distinct vacua, corresponding to $n=0,1,\dots,9$.\\
\indent First, consider that the quark fields transform under the representation assignment of Case A (Eq.~(\ref{qnp222})), namely the first quark mass model. The mass matrices for the down-type and up-type quarks are then
\begin{equation}
M_{d}=\frac{1}{\sqrt{2}}\begin{pmatrix}
 y_{2}^dv_3& 0&y_{1}^dv_1e^{i(\theta+\frac{n\pi}{5})}\\
0&y_{2}^dv_3e^{\frac{3in\pi}{5}}& y_{1}^dv_1e^{i\theta}\\
y_{3}^dv_1e^{i(\theta+\frac{n\pi}{5})}& y_{3}^dv_1e^{i\theta}& 0
\end{pmatrix},
\end{equation}
\begin{equation}
M_{u}=\frac{1}{\sqrt{2}}\begin{pmatrix}
 y_{2}^uv_3e^{-\frac{3in\pi}{5}}& 0&y_{1}^uv_1e^{-i\theta}\\
0&y_{2}^uv_3& y_{1}^uv_1e^{-i(\theta+\frac{n\pi}{5})}\\
y_{3}^uv_1e^{-i\theta}& y_{3}^uv_1e^{-i(\theta+\frac{n\pi}{5})}& 0
\end{pmatrix}.
\end{equation}
Note that within the framework of spontaneous CP violation, the quark Yukawa coupling coefficients are real. The mass-squared matrices for the down-type and up-type quarks are then, respectively:
\begin{equation}
M_d M_d^\dagger = \frac{1}{2}
\begin{pmatrix}
y_{1}^2 v_1^2 + y_{2}^2 v_3^2 & y_{1}^2 v_1^2 e^{\frac{in\pi}{5}} & y_{2} y_{3} v_1 v_3 e^{-i(\theta+\frac{n\pi}{5})} \\
y_{1}^2 v_1^2 e^{-\frac{in\pi}{5}} & y_{1}^2 v_1^2 + y_{2}^2 v_3^2 & y_{2} y_{3} v_1 v_3 e^{-i\left(\theta-\frac{3n\pi}{5}\right)} \\
y_{2} y_{3} v_1 v_3 e^{i(\theta+\frac{n\pi}{5})} & y_{2} y_{3} v_1 v_3 e^{i\left(\theta-\frac{3n\pi}{5}\right)} & 2 y_{3}^2 v_1^2
\end{pmatrix},
\end{equation}
\begin{equation}
M_u M_u^\dagger = \frac{1}{2}
\begin{pmatrix}
y_{1}^2 v_1^2 + y_{2}^2 v_3^2 & y_{1}^2 v_1^2 e^{i\frac{n\pi}{5}} & y_{2} y_{3} v_1 v_3 e^{i\left(\theta - \frac{3n\pi}{5}\right)} \\
y_{1}^2 v_1^2 e^{-\frac{in\pi}{5}} & y_{1}^2 v_1^2 + y_{2}^2 v_3^2 & y_{2} y_{3} v_1 v_3 e^{i\left(\theta + \frac{n\pi}{5}\right)} \\
y_{2} y_{3} v_1 v_3 e^{-i\left(\theta-\frac{3n\pi}{5}\right)} & y_{2} y_{3} v_1 v_3 e^{-i\left(\theta + \frac{n\pi}{5}\right)} & 2 y_{3}^2 v_1^2
\end{pmatrix}.
\end{equation}
Note that, for notational simplicity, the superscripts distinguishing down-type and up-type quark Yukawa couplings have been omitted in the above expressions, i.e., $y_i = y_{i}^{(d,u)}$.
Their determinant structures are identical:
\begin{equation}
\det(M_d M_d^\dagger, M_u M_u^\dagger) = \begin{cases}
\qquad \quad 0, & n \text{ is odd},\\
\dfrac{1}{2} y_1^2 y_2^2 y_3^2 v_1^4 v_3^2, & n \text{ is even}.
\end{cases}
\end{equation}
Based on the above Hermitian matrices and determinants, in order to realize the experimentally observed non-zero quark masses and a non-block-diagonal CKM matrix, the Yukawa couplings must strictly satisfy the following conditions:
\begin{equation}
y_1 \neq 0,\quad y_2 \neq 0,\quad y_3 \neq 0,\quad n \text{ is a non-zero even integer}.
\end{equation}
To prevent this vacuum from degenerating into the vacuum C-N-3b, $n$ must not be divisible by 5. In this case, $n$ can take four different even values, which can be labeled as $n=2,4,6,8$.\\
\indent Further investigation reveals that the above Hermitian matrices $M_d M_d^\dagger$ and $M_u M_u^\dagger$ can be simultaneously brought into block-diagonal form by the same unitary transformation, which in turn leads to a block-diagonal CKM mixing matrix. The detailed analysis is as follows:\\
\indent For brevity, we define:
\begin{equation}
\begin{aligned}
a_d = y_{2}^dv_3,\quad b_d = y_{1}^dv_1,\quad c_d = y_{3}^dv_1,\\
a_u = y_{2}^uv_3,\quad b_u = y_{1}^uv_1,\quad c_u = y_{3}^uv_1.
\end{aligned}
\end{equation}
One can find a unitary transformation matrix:
\begin{equation}
\label{U0CKM}
U_0 = \frac{1}{\sqrt{2}}\begin{pmatrix}
1 & 1 & 0 \\
e^{-\frac{in\pi}{5}} & -e^{-\frac{in\pi}{5}} & 0 \\
0 & 0 & \sqrt{2}
\end{pmatrix},
\end{equation}
which satisfies the unitarity condition $U_0^\dagger U_0=I_3$, and it can simultaneously block-diagonalize the mass-squared matrices of the down-type and up-type quarks. Consider first the down-type quarks. In the $U_0$ basis, $H_d = M_d M_d^\dagger$ is brought into block-diagonal form, i.e., $H_d'$ becomes block-diagonal:
\begin{equation}
H_d'=U_0^\dagger H_d U_0 = 
\frac12
\begin{pmatrix}
a_d^2 + 2b_d^2 & 0 & \sqrt{2} a_d c_d e^{-i\varphi_d} \\
0 & a_d^2 & 0 \\
\sqrt{2} a_d c_d e^{i\varphi_d} & 0 & 2c_d^2
\end{pmatrix},
\end{equation}
where the phase angle $\varphi_d$ satisfies
\begin{equation}
 \varphi_d = \theta + \frac{n\pi}{5}, \qquad n=2,4,6,8.
\end{equation}
Its three eigenvalues can be obtained as:
\begin{equation}
\begin{aligned}
&\lambda_{1}^d = \frac{a_d^2}{2},\\& \lambda_{\pm}^d = \frac{a_d^2 + 2b_d^2 + 2c_d^2 \pm \sqrt{(a_d^2+2b_d^2-2c_d^2)^2 + 8a_d^2c_d^2}}{4}.
\end{aligned}
\end{equation}
One sees that these eigenvalues are independent of the phase angle $\varphi_d$. Sorted in ascending order, they correspond respectively to the squared masses of the three down-type quarks $d,s,b$. One may assume that $\lambda_{1}^d$ corresponds to the lightest $d$ quark, while $\lambda_{-}^d$ and $\lambda_{+}^d$ correspond to the $s$ and $b$ quarks, respectively, i.e.,
\begin{equation}
m_{d}^2 = \lambda_{1}^d,\quad m_{s}^2 = \lambda_{-}^d,\quad m_{b}^2 = \lambda_{+}^d.
\end{equation}
Hence the experimentally measured quark masses $m_{d,s,b}$ can be used to solve for $a_d,b_d,c_d$.
Furthermore, the normalized eigenvectors corresponding to these eigenvalues are:
\begin{equation}
\lambda_{1}^d: \begin{pmatrix} 0 \\ 1 \\ 0 \end{pmatrix},\qquad
\lambda_{-}^d: \begin{pmatrix} -\sin\alpha_d\, e^{-i\varphi_d}\\ 0 \\ \cos\alpha_d \end{pmatrix}, \qquad \lambda_{+}^d: \begin{pmatrix} \cos\alpha_d \\ 0 \\ 
\sin\alpha_d \, e^{i\varphi_d}\end{pmatrix},
\end{equation}
where the mixing angle $\alpha_d$ satisfies
\begin{equation}
\tan(2\alpha_d) = \frac{2\sqrt{2}\,a_d c_d}{a_d^2+2b_d^2-2c_d^2}.
\end{equation}
Thus, for the down-type quarks, the unitary matrix from the original quark basis to the mass eigenbasis is
\begin{equation}
U_{d_L} = U_0 \cdot V_d,
\end{equation}
with
\begin{equation}
V_d = \begin{pmatrix}
0 &  -\sin\alpha_d\, e^{-i\varphi_d} &\cos\alpha_d\\
1 & 0 & 0 \\
0  &\cos\alpha_d& \sin\alpha_d\,e^{i\varphi_d}
\end{pmatrix}.
\end{equation}
The diagonalization for the up-type quarks proceeds similarly. In the $U_0$ basis, $H_u = M_u M_u^\dagger$ is brought into block-diagonal form, i.e., $H_u'$ becomes block-diagonal:
\begin{equation}
H_u' = U_0^\dagger H_u U_0 = \frac{1}{2}
\begin{pmatrix}
a_u^2 + 2b_u^2 & 0 & \sqrt{2}\,a_uc_u\, e^{-i\varphi_u} \\
0 & a_u^2 & 0 \\
\sqrt{2}\,a_uc_u\, e^{i\varphi_u} & 0 & 2c_u^2
\end{pmatrix},
\end{equation}
where the phase angle $\varphi_u$ satisfies
\begin{equation}
\varphi_u = -\theta + \frac{3n\pi}{5}, \qquad n=2,4,6,8.
\end{equation}
The three eigenvalues are found to be:
\begin{equation}
\begin{aligned}
&\lambda_{1}^u = \frac{a_u^2}{2},\\&
\lambda_{\pm}^u = \frac{a_u^2 + 2b_u^2 + 2c_u^2 \pm\sqrt{(a_u^2+2b_u^2-2c_u^2)^2 + 8a_u^2c_u^2}}{4},
\end{aligned}
\end{equation}
and again these eigenvalues are independent of the phase angle $\varphi_u$. Sorted in ascending order, they correspond respectively to the squared masses of the three up-type quarks $u,c,t$. One may assume that $\lambda_{1}^u$ corresponds to the lightest $u$ quark, while $\lambda_{-}^u$ and $\lambda_{+}^u$ correspond to the $c$ and $t$ quarks, respectively, i.e.,
\begin{equation}
m_{u}^2 = \lambda_{1}^u,\quad m_{c}^2 = \lambda_{-}^u,\quad m_{t}^2 = \lambda_{+}^u.
\end{equation}
Hence the experimentally measured up-type quark masses $m_{u,c,t}$ can be used to solve for $a_u,b_u,c_u$.
The normalized eigenvectors corresponding to these eigenvalues are:
\begin{equation}
\lambda_{1}^u: \begin{pmatrix} 0 \\ 1 \\ 0 \end{pmatrix},\qquad
\lambda_{-}^u: \begin{pmatrix} -\sin\alpha_u\, e^{-i\varphi_u}\\ 0 \\ \cos\alpha_u \end{pmatrix}, \qquad \lambda_{+}^u: \begin{pmatrix} \cos\alpha_u \\ 0 \\ 
\sin\alpha_u \, e^{i\varphi_u}\end{pmatrix},
\end{equation}
where the mixing angle $\alpha_u$ satisfies
\begin{equation}
\tan(2\alpha_u) = \frac{2\sqrt{2}\,a_u c_u}{a_u^2+2b_u^2-2c_u^2}.
\end{equation}
Therefore, for the up-type quarks, the unitary matrix from the original quark basis to the mass eigenbasis is
\begin{equation}
U_{u_L} = U_0 \cdot V_u,
\end{equation}
with
\begin{equation}
V_u = \begin{pmatrix}
0 &  -\sin\alpha_u\, e^{-i\varphi_u} &\cos\alpha_u\\
1 & 0 & 0 \\
0  &\cos\alpha_u& \sin\alpha_u\,e^{i\varphi_u}
\end{pmatrix}.
\end{equation}
Now we obtain the analytical form of the CKM matrix. Using the unitarity of $U_0$, the CKM matrix simplifies to
\begin{equation}
V_{\text{CKM}} = U_{u_L}^\dagger U_{d_L} = V_u^\dagger V_0^\dagger V_0 V_d = V_u^\dagger V_d.
\end{equation}
A direct calculation yields
\begin{equation}
\setlength{\arraycolsep}{2.6pt}
V_u^\dagger V_d = \scalebox{0.96}{$\begin{pmatrix}
1 & 0 & 0 \\
0 & \sin\alpha_u \sin\alpha_d \, e^{i(\varphi_u - \varphi_d)} + \cos\alpha_u \cos\alpha_d & 
 \cos\alpha_u \sin\alpha_d \, e^{i\varphi_d}-\sin\alpha_u \cos\alpha_d \, e^{i\varphi_u} 
\\
0 & 
 \sin\alpha_u \cos\alpha_d \, e^{-i\varphi_u} -\cos\alpha_u \sin\alpha_d \, e^{-i\varphi_d} 
& 
\cos\alpha_u \cos\alpha_d + \sin\alpha_u \sin\alpha_d \, e^{i(\varphi_d - \varphi_u)}
\end{pmatrix}.$}
\end{equation}
One can see that the CKM matrix takes a block-diagonal form, with $V_{us} = V_{ub} = V_{cd} = V_{td} = 0$. This means that the first generation quarks ($u$ and $d$) are decoupled from the second and third generation quarks, producing no cross-generational mixing. Only mixing between the second and third generations ($c, t$ and $s, b$) is possible. However, experimentally all off-diagonal elements of the CKM matrix are non-zero, so the above form is in severe contradiction with experimental observations.\\
\indent Specifically, $M_d M_d^\dagger$ and $M_u M_u^\dagger$ possess a highly symmetric structure: their $2 \times 2$ sub-matrices formed by the first two rows and first two columns are
\begin{equation}
\label{dduuscpv}
\frac12 \begin{pmatrix}
y_1^2 v_1^2 + y_2^2 v_3^2 & y_1^2 v_1^2 e^{i n\pi/5} \\
y_1^2 v_1^2 e^{-i n\pi/5} & y_1^2 v_1^2 + y_2^2 v_3^2
\end{pmatrix},
\end{equation}
where the off-diagonal entries carry the same phase factor $e^{\pm i n\pi/5}$. The elements in the third row and third column are all $y_3^2 v_1^2$. The crucial differences appear in the $(1,3)$ and $(2,3)$ matrix elements (and their complex conjugates), which carry different phases in $M_d M_d^\dagger$ and $M_u M_u^\dagger$:
\begin{equation}
\label{dduu1323scpv}
\begin{aligned}
(M_d M_d^\dagger)_{13} &\propto e^{-i(\theta + n\pi/5)}, \quad (M_d M_d^\dagger)_{23} \propto e^{-i(\theta - 3n\pi/5)}, \\
(M_u M_u^\dagger)_{13} &\propto e^{i(\theta - 3n\pi/5)}, \quad (M_u M_u^\dagger)_{23} \propto e^{i(\theta + n\pi/5)}.
\end{aligned}
\end{equation}
However, there exists a unitary matrix $U_0$ (Eq.~\ref{U0CKM}) that can exactly block-diagonalize both matrices simultaneously, i.e., $U_0^\dagger M_d M_d^\dagger U_0$ and $U_0^\dagger M_u M_u^\dagger U_0$ both take a $(2+1)$ block-diagonal form. This result stems from the following two points:
\begin{itemize}
\item[(1)] The first two columns of $U_0$ provide a set of orthogonal basis vectors for the first two-dimensional subspace that diagonalize that $2 \times 2$ sub-block;
\item[(2)] Although the $(1,3)$ and $(2,3)$ entries are individually non-zero, under the $U_0$ transformation they are recombined due to phase matching relations and precisely fall into different blocks, thereby producing no cross-block mixing.
\end{itemize}
This possibility of simultaneous block diagonalization originates from the $D_5$ symmetry imposed by the model, which forces the magnitudes of the complex vacua that can realize spontaneous CP violation to satisfy $|v_1| = |v_2|$, $|v_3| = |v_4|$, and consequently makes the mass-squared matrices of the up- and down-type quarks possess a highly consistent structure in the first two dimensions. It is for this reason that both can be simultaneously block-diagonalized by the same $U_0$, ultimately leading to a CKM matrix of block-diagonal form, i.e., the quark mixing between different blocks is strictly zero.

\subsection{C-N-3a Model II}

\indent Now consider the C-N-3a Model II, i.e., the case where the quark fields transform under the representation assignment of Case B (Eq.~(\ref{qnp22'2'})). The mass matrices for the down-type and up-type quarks are then
\begin{equation}
M_d = \frac{1}{\sqrt{2}} \begin{pmatrix}
y_{3}^d v_3 e^{\frac{3in\pi}{5}} & y_{1}^d v_1 e^{i\theta} & y_{2}^d v_1 e^{i(\theta+\frac{n\pi}{5})} \\
y_{1}^d v_1 e^{i(\theta+\frac{n\pi}{5})} & y_{3}^d v_3 & y_{2}^d v_1 e^{i\theta} \\
y^{d}_{4} v_3 & y^{d}_{4} v_3 e^{\frac{3in\pi}{5}} & 0
\end{pmatrix},
\end{equation}
\begin{equation}
M_u = \frac{1}{\sqrt{2}} \begin{pmatrix}
y_{3}^u v_3 & y_{1}^u v_1 e^{-i(\theta+\frac{n\pi}{5})} & y_{2}^u v_1 e^{-i\theta} \\
y_{1}^u v_1 e^{-i\theta} & y_{3}^u v_3 e^{-\frac{3in\pi}{5}} & y_{2}^u v_1 e^{-i(\theta+\frac{n\pi}{5})} \\
y_{4}^u v_3 e^{-\frac{3in\pi}{5}} & y_{4}^u v_3 & 0
\end{pmatrix}.
\end{equation}
Their Hermitian matrices are:
\begin{equation}
\begin{aligned}
M_d M_d^\dagger = \frac{1}{2}
\begin{pmatrix}
A_d & B_d & C_d\\
B_d^* & A_d & D_d \\
C_d^* & D_d^* & E_d
\end{pmatrix},
\end{aligned}
\end{equation}
\begin{equation}
\begin{aligned}
M_u M_u^\dagger = \frac{1}{2}
\begin{pmatrix}
A_u & B_u & D_u^*\\
B_u^* & A_u & C_u^* \\
D_u & C_u & E_u
\end{pmatrix},
\end{aligned}
\end{equation}
where
\begin{equation}
\begin{aligned}
A_{d,u} &=  (y_1^2 + y_2^2)v_1^2 + y_3^2 v_3^2, \\
B_{d,u} &= \left(2 y_1 y_3 v_1 v_3 \cos\left(\theta-\tfrac{n\pi}{5}\right) + y_2^2 v_1^2\right) e^{-\frac{in\pi}{5}} , \\
C_{d,u} &= y_1 y_4 v_1 v_3 e^{i\left(\theta-\frac{3n\pi}{5}\right)} + y_3 y_4 v_3^2 e^{\frac{3in\pi}{5}}, \\
D_{d,u} &= y_1 y_4 v_1 v_3 e^{i\left(\theta+\frac{n\pi}{5}\right)} + y_3 y_4 v_3^2 e^{-\frac{3in\pi}{5}} , \\
E_{d,u} &= 2y_4^2v_3^2.
\end{aligned}
\end{equation}
Note that, for notational simplicity, the superscripts distinguishing down-type and up-type quark Yukawa couplings have been omitted in the above expressions, i.e., $y_i = y_{i}^{(d,u)}$. Their determinant structures are identical:
\begin{equation}
\det(M_d M_d^\dagger, M_u M_u^\dagger) = \begin{cases}
\qquad \qquad\qquad \quad0, & n\text{ is odd},\\
\dfrac{1+(-1)^n}{2} \cdot \dfrac{y_2^2 y_4^2v_1^2 v_3^2}{2} \left| y_1v_1 - y_3v_3 e^{i(\theta - \frac{n\pi}{5})} \right|^2 ,& n\text{ is even}.
\end{cases}
\end{equation}
Based on the above Hermitian matrices and determinants, in order for the CKM matrix to be non-block-diagonal and for the quark masses to be non-zero, the Yukawa couplings must strictly satisfy the following conditions:
\begin{equation}
 y_2 \neq 0,\quad \quad y_4 \neq 0, \quad \quad y_1 v_1 \neq y_3 v_3 e^{i(\theta - \frac{n\pi}{5})}, \quad n \text{ is a non-zero even integer}.
\end{equation}
Similar to C-N-3a Model I, this model also possesses a unitary transformation matrix $U_0$ (Eq.~\ref{U0CKM}) that simultaneously block-diagonalizes $M_d M_d^\dagger$ and $M_u M_u^\dagger$, resulting in a block-diagonal CKM matrix, and should therefore be excluded as well. 
In fact, by examining every vacuum listed in Table~\ref{scpvcvevs} under both quark models in the quark sector and under all possible permutation assignments of the representations in Eq.~(\ref{D52plus1}), one finds that, due to the presence of this unitary matrix $U_0$ (Eq.~\ref{U0CKM}), they all ultimately lead to a block-diagonal CKM mixing matrix. Hence, in the $D_5$ 4HDM, all vacua that admit SCPV are inconsistent with experimental observations.

\section{Conclusion}
\label{Sec:Conclusion}

Based on the complex vacua that can lead to spontaneous CP violation obtained in the scalar sector of the $D_5$ 4HDM, we further extend them to the quark sector. When considering spontaneous CP violation, we require all parameters in the Lagrangian to be real, i.e., only the scalar interactions are allowed to undergo spontaneous CP violation. In the Yukawa sector, the Yukawa interactions are constructed in a $D_5$-invariant manner. Only when the three generations of fermions are assigned as combinations of singlet and doublet representations under the $D_5$ symmetry can the experimentally required non-zero quark masses and non-block-diagonal CKM mixing matrix be satisfied. On this basis, all possible quark fields can be unified into two distinct quark mass models.

In the framework of spontaneous CP violation, due to the strong constraints of the $D_5$ symmetry, special relations among the moduli of the vacuum expectation values are enforced, so that the mass-squared matrices of the down-type and up-type quarks (i.e., $M_d M_d^\dagger$ and $M_u M_u^\dagger$) can be simultaneously block-diagonalized by the same unitary transformation, thereby forcing the CKM matrix to take a block-diagonal form and making it impossible to generate non-zero mixing angles and CP violation. Within this framework, the only two types of quark mass models both exhibit this behavior in all complex vacua that can realize SCPV. This result rules out the possibility of the model explaining experimental observations under SCPV.

Thus we have demonstrated that in $D_5$ 4HDM, if one relies solely on spontaneous CP violation, the quark sector cannot produce a CKM matrix consistent with experimental requirements, and consequently cannot account for the observed CP violation. For the $D_5$ 4HDM, a flavor mixing and CP violation structure consistent with experiment can only be realized within the framework of explicit CP violation.

\section*{Acknowledgments}
This work is supported by the Fundamental Research Funds for the Central Universities, the One Hundred Talent Program of Sun Yat-sen University, China, and the Guangdong Natural Science Foundation (Project No. 2026A1515012641).
\appendix

\end{document}